\documentclass[%
 superscriptaddress,
 preprint,
 amsmath,amssymb,
 notitlepage,
 floatfix
]{revtex4-2}

\usepackage{graphicx}% Include figure files
\usepackage{dcolumn}% Align table columns on decimal point
\usepackage{bm}% bold math
\usepackage{verbatim}
\usepackage{color}
\usepackage[caption=false]{subfig}
\usepackage{tabularx}
\usepackage[section]{placeins}
\begin{document}
\title{Probing Exotic Phases Via Stochastic Gravitational Wave Spectra}
\author{Joshua Berger}
\affiliation{Colorado State University, Fort Collins, Colorado, 80523, USA}
\author{Amit Bhoonah}
\affiliation{Department of Physics and Astronomy and IQ Initiative,  University of Pittsburgh, Pittsburgh, PA 15260, USA}
\author{Biswajit Padhi}
\affiliation{Colorado State University, Fort Collins, Colorado, 80523, USA}

\date{\today}

\begin{abstract}
Stochastic backgrounds of gravitational waves (GWs) from the pre-BBN era offer a unique opportunity to probe the universe beyond what has already been achieved with the Cosmic Microwave Background (CMB). If the source is short in duration, the low frequency tail of the resulting GW spectrum follows a universal frequency scaling dependent on the equation of state of the universe when modes enter the horizon. We demonstrate that the distortion of the equation of state due to massive particles becoming non-relativistic can lead to an observable dip in the GW spectrum. To illustrate this effect, we consider a first order chiral symmetry breaking phase transition in the weak-confined Standard Model (WCSM).  The model features a large number of pions and mostly elementary fermions with masses just below the critical temperature for the phase transition. 
These states lead to a 20$\%$ dip in the GW power.  We find potential sensitivity to the distortions in the spectrum to future GW detectors such as LISA, DECIGO, BBO, and $\mu$Ares.
\end{abstract}

\maketitle
\newpage

\section{Introduction}

The prospects for gravitational waves (GWs) to probe exotic particle phenomena have gained increasing attention in light of their discovery by the LIGO scientific collaboration \cite{LIGOScientific:2016aoc}. The proof of principle that they provided has led to the proposal for a large number of new GW telescopes covering a wide range of frequencies, from (10-100) nHz with Pulsar Timing Arrays (PTAs) \cite{Manchester:2006xj,Kramer:2013kea,NANOGrav:2020bcs} to the mHz to 10 kHz window using ground and space based telescopes \cite{Kawamura:2006up,Harry:2006fi,TianQin:2020hid}. The range between PTAs and telescopes is challenging to cover, but there exist proposals such as the $\mu$Ares detector \cite{Sesana:2019vho} and through the use of asteroids \cite{Fedderke:2021kuy}. At much higher frequencies, in the MHz-GHz range, novel laboratory techniques have already constrained gravitational wave backgrounds \cite{Holometer:2016qoh,Akutsu:2008qv} and there are plans for future detectors capable of extending current sensitivity further \cite{Ito:2022rxn,Berlin:2023grv}. These observatories will significantly broaden the scope of new physics that can be explored through gravitational waves in the coming decades.

This development has the potential to radically alter our understanding of the universe in much the way detailed studies of the Cosmic Microwave Background (CMB) \cite{1965ApJ...142..419P} have over the past few decades.  In studying the CMB \cite{COBE:1992syq,WMAP:2012fli,Planck:2018nkj,Swetz:2010fy}, we were able to infer with great precision the properties of the early universe fluid by studying the spectrum of background photons, as well as the anisotropies therein. It would be remarkable indeed if the same level of detailed study could be performed by GW experiments. For the purposes of understanding the early universe, it is particularly interesting to learn what can be inferred from GWs emitted prior to big bang nucleosynthesis (BBN) \cite{1967ApJ...148....3W}. The CMB can only directly probe the universe back to recombination at $T \sim \text{eV}$, but observations of the primordial abundance of light elements probes the universe back to $T \sim \text{MeV}$. Above this, we thus far have no real probes and the best hope for accessing this era is in GWs; the universe is essentially transparent to them since they only interact via gravity and this makes them ideal messengers for probing the pre-BBN era. This is, however, not without challenges because the universe was really tiny at these times and effects like the Sachs-Wolfe effect \cite{1967ApJ...147...73S,1968Natur.217..511R}, that have proven essential to determining our picture of the universe from CMB photons, are suppressed.

Several early universe events, such as first order phase transitions \cite{Turner:1990rc,1986MNRAS.218..629H,Kamionkowski:1993fg}, hybrid preheating after inflation \cite{Garcia-Bellido:2007nns}, dynamics of topological defects \cite{Vilenkin:1981bx}, and soliton collisions \cite{Helfer:2018vtq}, can produce GWs in the pre-BBN era.  These are referred to as ``stochastic'' and would constitute a nearly isotropic background of gravitational radiation just like the CMB is a nearly isotropic background of electromagnetic radiation. In this work, we focus for concreteness on GWs produced via a first order phase transition, though none of the qualitative results of our analysis rely on much beyond assuming that the GWs are produced nearly instantaneously and isotropically on cosmological times. 

A first order phase transition proceeds by nucleation of the new vacuum phase as bubbles that expand. Eventually, those bubbles collide violently enough to provide a first source of GWs. Colliding bubbles can also percolate and interact with the surrounding thermal plasma, causing bulk motion in the fluid which, in turn, provides two further sources of GWs: acoustic (sound wave) and magnetohydrodynamic disturbances \cite{Caprini:2009fx}. The spectrum of GW emission from these processes have been simulated numerically and depends on multiple parameters such as the duration of the phase transition, the amount of latent heat released, and the fraction of energy which gets converted into the bulk motion of the plasma (see, for example, Refs. \cite{Caprini:2009fx,Caprini:2015zlo,Caprini:2019egz} for more detailed discussions). The precise spectrum of gravitational waves from these three sources is an area of ongoing research. 

Despite these model dependent features, a pair of related recent papers~\cite{Hook:2020phx,Brzeminski:2022haa} pointed out the interesting fact that the low frequency tail of GWs from any ``instantaneous'' event would have a universal spectrum which only depends on features of the universe when those modes enter the horizon. Intriguingly, the spectrum is sensitive to the equation of state of the plasma. Provided the universe is in a radiation dominated phase, as expected in a standard cosmology, we expect $w \approx 1/3$. Nevertheless, there can be small deviations from $w = 1/3$ and a careful measurement of the spectrum of GWs would be sensitive to such deviations. The equation of state has proven to be a useful probe of early universe phases, as illustrated in e.g.\ \cite{Chung:2011hv,Chung:2011it}.

There are several potential sources such a distortion, but all require a breaking of scaling invariance.  Any scale-invariant theory will have $w = 1/3$ exactly as
\begin{equation}
\partial_\mu J^\mu_\lambda = T^\mu_\mu = \rho - 3 \, p,
\end{equation}
where $J_\lambda$ is the scale transformation current. One could imagine several small sources of breaking in the radiation-dominated era. The recent works metioned above, Refs.~\cite{Hook:2020phx,Brzeminski:2022haa}, considered interactions inducing a scale anomaly, leading to $w - 1/3 \propto \beta$, where $\beta$ is the beta function of the theory, and free-streaming of particles that decouple from the plasma.

In this work, we focus primarily on scale-invariance breaking due to the masses of particles. We demonstrate that when several degrees of freedom simultaneously become non-relativistic, the resulting distortion of the equation of state is sufficiently significant to be visible in future GW observations. Since there are hundreds of bosonic degrees of freedom at high temperatures, the resulting distortion goes like $g / g_*$, where $g$ is the number of degrees of freedom that are becoming non-relativistic and $g_*$ is the total number of relativistic degrees of freedom. Unlike the signals considered previously, this distortion is transient.  If the mass of the particles is not too far from the temperature at which the GWs are sourced, then a localized distortion of the long-wavelength tail of the GW spectrum could be observed.

To demonstrate how this could work and could help determine qualitative information about the non-standard early universe cosmology, we consider the weak-confined standard model (WCSM) presented in Ref.~\cite{Berger:2019yxb,Lohitsiri:2019wpq}. In that model, a scalar field coupling to the $SU(2)_L$ gauge kinetic operator leads to an effective shift in the scalar field coupling. This shift occurs when the scalar field has a non-zero expectation value, a mechanism first presented in Ref.~\cite{Ipek:2018lhm}. If the shift is toward a stronger $SU(2)_L$ coupling, then it is possible for the weak force to become strongly coupled at $T > T_c$, the critical temperature for the electroweak crossover of standard cosmology. In a strongly coupled $SU(2)_L$ phase, $B+L$-violating transitions would occur readily, opening the door for a possible baryogenesis mechanism. For such a mechanism to work, the standard cosmology phase would need to be restored at $T < T_c$ so that the electroweak crossover does not washout the generated asymmetry.

More to the point of this work, the transition to strong coupling could be a first order phase transition, leading naturally to a source of stochastic gravitational waves. What makes the WCSM a particularly interesting source is that the nature of the phase of strongly-coupled weak force is in question. It is unclear whether this would be a chiral symmetry-breaking confined phase, as with standard model Quantum Chromodynamics (QCD), or an infrared conformal phase. In the latter case, the only distortions away from $w = 1/3$ would be due to the scale anomaly of the remaining interactions (QCD, electromagnetic, and Yukawa). On the other hand, the former would break scale invariance more strongly. A number of species would have masses near and a bit below the critical temperature, leading to modifications of the equation of state in the region below the phase transition. We show that in a chiral symmetry-breaking phase, pions becoming non-relativistic below the phase transition lead to more than 10\% deviation from $w = 1/3$, correspondingly leading to an order 10\% dip in the characteristic strain or 20\% dip in the GW power at a frequency corresponding to modes that enter the horizon just as the pions are becoming non-relativistic. For a WCSM phase transition above the would-be electroweak crossover, such a dip could yield potentially observable gravitational wave signatures.

In this work, we develop a calculation of the distortion of the gravitational waves due to particle masses in the WCSM. This requires modifying the results of Refs.~\cite{Hook:2020phx,Brzeminski:2022haa} to account for a time-dependent equation of state. We solve the resulting gravitational wave equations numerically for benchmark models.  We find potential sensitivity to the distortions in the spectrum to future GW detectors such as the Laser Interferometer Space Antenna (LISA)  \cite{Amaro-Seoane:2012vvq}, Deci-hertz Interferometer Gravitational wave Observatory (DECIGO) \cite{Kawamura:2006up}, Big Bang Observer (BBO) \cite{Harry:2006fi}, and the $\mu$Ares detector. 

The remainder of this paper is structured as follows. In Section \ref{sec:formalism}, we develop the formalism required to study time-dependent distortions of the equation of state and demonstrate the effect of a small number of non-relativistic species on the gravitational wave spectrum. We then review the relevant details of the WCSM model in Section \ref{sec:model}. The results of our study are presented in Section \ref{sec:results}. Finally, in Section \ref{sec:outlook}, we discuss the implications of our results for future measurements.

\section{General Formalism}\label{sec:formalism}

\subsection{Gravitational Wave Equation}

Our starting point is to consider a Friedmann-Robertson-Lema\^{i}tre-Walker universe with metric
\begin{equation}
ds^2 = dt^2 - a(t)^2 \, d\mathbf{x}^2.
\end{equation}
Following Refs.~\cite{Hook:2020phx,Brzeminski:2022haa}, we work in conformal time $\tau = \int dt/a(t)$ and expand the FLRW metric in small perturbations to obtain the linearized gravitational wave equation in comomving momentum space:
\begin{equation}\label{eq:GWSourceFree}
h^{\prime\prime} + 2 \, \mathcal{H} \, h^\prime + k^2 \, h = 0,
\end{equation}
where $\prime$ denotes derivatives with respect to conformal time, $\mathcal{H} = a^\prime / a$ is the conformal Hubble rate, $k$ is the comoving wavenumber, and $h$ is the metric perturbation amplitude.

Suppose that at a time $\tau_{*}$ an instantaneous event  (meaning that the duration of the event $\Delta \tau_{*} \ll 1/\mathcal{H}_{*}$, where $\mathcal{H}^{*} \equiv \mathcal{H}(\tau_{\star})$)  generates gravitational waves. Then one must add a source on the right-hand side of~\eqref{eq:GWSourceFree}, and the resulting equation can be solved using Green's function methods, which is equivalent to solving the source-free equation with initial conditions
\begin{equation}
h(\tau_*) = 0, \qquad h^\prime(\tau_*) = J_*,
\end{equation}
where $J_*$ is related to the dimensionless anisotropic stress inducing the gravitational waves,
\begin{equation}
J_* = a^2 \, \frac{32 \, \pi \, G \, \rho}{3} \, \epsilon^{ij} \, \Pi_{ij},
\end{equation}
projecting onto the GW polarization tensor $\epsilon$.

When $k \ll \mathcal{H}_*$, the modes are initially frozen outside the horizon. As they enter the horizon, they become sensitive the equation of state determining $\mathcal{H}$. When $k \gg \mathcal{H}$, the modes lose sensitivity to the equation of state and simply oscillate. Thus, if a number of degrees of freedom are becoming non-relativistic at the time when a given mode is entering the horizon, the spectrum for that mode gets distorted. 

We demonstrate that future experiments should have sensitivity to this distortion. To see this, we must first generalize the calculation of \cite{Hook:2020phx,Brzeminski:2022haa}. In that work, the authors considered the equation of state to be approximately constant. For the effect we are interested in, this approximation is not accurate as the modification to the equation of state is happening on a Hubble time, which is the same as the time scale for the modes to enter the horizon. 

Defining the equation of state $w = p/\rho$ and plugging the first Friedmann equation $\mathcal{H}^2 = 8 \, \pi \, G \, \rho \, a^2/3$ into the acceleration Friedmann equation, we get an equation for $a$ in terms of $w$:
\begin{equation}
2 \, a \, a^{\prime\prime} = (1 - 3 \, w) \, (a^\prime)^2.
\end{equation}
This equation can be solved to determine $\mathcal{H}$ as
\begin{equation}
\mathcal{H} = \frac{2}{\tau + 3 \, \int_0^\tau w(\tau_1) \, d\tau_1}.
\end{equation}
As $w(\tau)$ can be determined from the matter content of the universe, we can solve the wave equation~\eqref{eq:GWSourceFree}, at least numerically.

It will be helpful to change variables in equation~\eqref{eq:GWSourceFree}, defining
\begin{equation}
    \tilde{h} = \tilde{\tau} \, h, \qquad \tilde{\tau} = k\, \tau,\qquad \tilde{\mathcal{H}} = \frac{\mathcal{H}}{k}.
\end{equation}
In terms of these variables, equation~\eqref{eq:GWSourceFree} becomes
\begin{equation}\label{eq:gw-reformulation}
    \tilde{h}^{\prime\prime} + 2 \, \left(\tilde{\mathcal{H}} - \frac{1}{\tilde{\tau}}\right) \, \tilde{h}^\prime + \left(1 + \frac{2}{\tilde{\tau}} \, \left[\frac{1}{\tilde{\tau}} - \tilde{\mathcal{H}}\right]\right) \, \tilde{h} = 0,
\end{equation}
with primes now denoting derivatives with respect to $\tilde{\tau}$.  In terms of these variables, for a purely conformal equation of state $w = 1/3$, $\tilde{H} = 1/\tilde{\tau}$, and the equation reduces to that of a harmonic oscillator with angular frequency 1.

At these late times, the solution goes like
\begin{equation}
    h = A(k) \frac{J_* \tau_* \sin(k (\tau - \tau_*) + \phi)}{k \tau},
\end{equation}
where $\phi$ is a phase shift induced due to the change in equation of state and $A$ quantifies the change in amplitude relative to the case of no change in equation of state. The ensemble-averaged power is approximately a period-averaged power at this time, so that we can write
\begin{equation}
    \frac{dP}{dk} \approx \langle (h^\prime)^2 \rangle \approx \frac{A^2(k) J_*^2 \tau_*^2}{2 k \tau^2},
\end{equation}
independent of the phase shift. We can therefore examine $A^2(k)$ as an observable measure of the effect of the distortion of the equation of state.  

\subsection{Time-Dependent Equation of State}
\label{sec:eos}

Each species in equilibrium with temperature $T$ in the universe contributes an energy density
\begin{equation}
\rho_i = g_i \, \frac{T^4}{2\, \pi^2} \, \int_x^\infty dy \, \frac{y^2 \, \sqrt{y^2 - x^2}}{e^y \pm 1},
\end{equation}
and a pressure
\begin{equation}
p_i = g_i \, \frac{T^4}{6\, \pi^2} \, \int_x^\infty dy \, \frac{(y^2 - x^2)^{3/2}}{e^y \pm 1},
\end{equation}
where $+$ is for fermions and $-$ is for bosons, $g_i$ is the number of degrees of freedom corresponding to the species and $x = m_i /T$ for a species of mass $m_i$. The total energy density and pressure are obtained by simply summing over the species,
\begin{equation}
    \rho_{\text{tot}} = \sum_i \rho_i,\qquad p_{\text{tot}} = \sum_i p_i.
\end{equation}
From this, we can determine the equation of state parameter $w$ as a function of temperature via
\begin{equation}
    w(T) = \frac{p_{\text{tot}}}{\rho_{\text{tot}}}.
\end{equation}

\begin{figure}[h]
    \centering
    \includegraphics[width=0.7\textwidth]{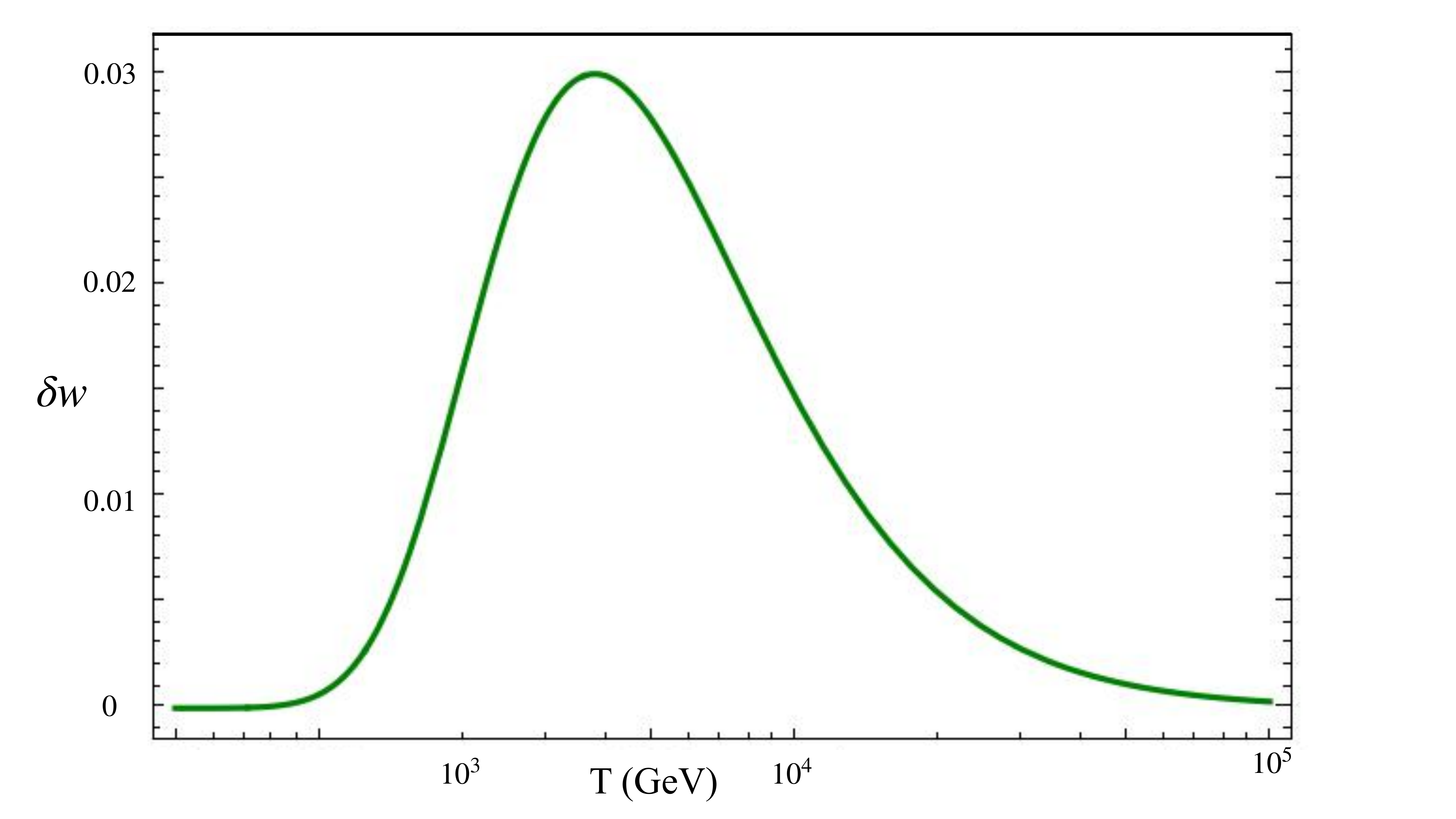}
    \caption{$\delta w$ for a thermal bath containing 2 bosonic species. The degrees of freedom with a mass of $10^4$ GeV decouple from the bath at temperatures of the order of their mass causing $w$ to deviate from 1/3.}
    \label{fig:deltaomegaillustration}
\end{figure}

Then, we can calculate $\delta w$ as a function of temperature using
\begin{equation}
    \delta w(T) = w(T) - \frac{1}{3}
\end{equation}
To illustrate the change in the equation of state, we first consider a toy model. We consider a scenario in which a thermal bath contains 2 different bosonic species. One of them has 40 relativistic degrees of freedom and the other species has 60 degees of freedom of mass $10^4$ GeV. %We use Eqs. (12)-(16) to calculate $\delta w(T)$. 
The behavior of $\delta w$ with respect to temperature in this toy scenario is shown in Figure~\ref{fig:deltaomegaillustration}. 
 In this toy model, a peak change in $w$ is found at $T = 0.38 m$ with $\delta w = 0.03$.  

To write $w$ as a function of $\tau$, we need the relation between conformal time and temperature, obtained by combining the first Friedmann equation with the conservation of entropy,
\begin{equation}
    \frac{d}{dT} (s \, a^3) = 0.
\end{equation}
We can then obtain the solution
\begin{equation}
    \tau(T) = \int_T^\infty \frac{ds(T^\prime)/dT^\prime}{2\, s(T^\prime) \, \sqrt{6 \, \pi G \, \rho_{\text{tot}}(T^\prime)}} \, dT^\prime.
\end{equation}
The entropy is given by the usual Euler relation
\begin{equation}
    s = \frac{\rho_{\text{tot}} + p_{\text{tot}}}{T}.
\end{equation}
Eqs. (18) and (19) allow us to express the conformal time as a function of temperature. This relation can be inverted to get the temperature as a function of conformal time, which allows us to express $\delta w$ as a function of $\tau$.
\subsection{Gravitational Wave Model}

The phase for the gravitational wave is arbitrary and depends on the random initial conditions of the universe. To estimate the sensitivity, we therefore consider the ensemble-averaged power in the gravitational waves,
\begin{equation}
    \rho_{\text{gw}} = \frac{1}{32 \, \pi \, G \, a^2} \, \sum_{\text{helicities}} \langle h^\prime_{ij}(\tau,\mathbf{x}) h^\prime_{ij}(\tau,\mathbf{x})\rangle
\end{equation}
In Fourier space, we define the power spectrum by
\begin{equation}
    \langle h(\tau,k)^2 \rangle = (2 \, \pi)^3 \, \delta^{(3)}(\mathbf{k} - \mathbf{k}^\prime) \, P_h(k).
\end{equation}
From this, we can define the differential power by
\begin{equation}
\frac{d\Omega_{\text{gw}}}{d\log k} = \frac{1}{\rho_c} \, \frac{k^5 \, P_h(\tau,k)}{2 \, (2 \, \pi)^3 \, a^2 \, G}.
\end{equation}

As discussed before, for a first order phase transition there are three contributions to the gravitational wave spectrum, from collisions of the bubbles themselves and from acoustic and magnetohydrodynamic distrubances induced by their expansion in the thermal plasma. The three sources are expected to combine linearly,
\begin{equation}
    h^{2}\Omega_{\text{gw}} = h^{2}\Omega_{\text{col}} + h^{2}\Omega_{\text{sw}} + h^{2}\Omega_{\text{MHD}}.
\end{equation}
To determine which one of these sources dominate we must consider the system undergoing a first order phase transition. In a strongly interacting theory such as the WCSM, one can reasonably expect that the bubble wall interacts sufficiently with the thermal plasma to cause bulk motion of the latter. In this scenario, numerous numerical studies (see for example \cite{Caprini:2015zlo,Hindmarsh:2015qta}) have found that sound waves are the dominant source of GWs compared to the collisions of bubbles themselves.\footnote{In the case of a supercooled electroweak phase transition, the dominant source comes from bubble wall collisions, but in our scenario, we expect the sound wave contribution to dominate.}  The peak frequency is given by \cite{Caprini:2015zlo,Caprini:2019egz}
\begin{equation}
    f_{\text{sw}} = 1.9 \times 10^{-5} \frac{1}{v_{w}}\left(\frac{\beta}{H_{*}}\right)\left(\frac{T_{*}}{100 \ \text{GeV}}\right)\left(\frac{g_{*}}{100}\right)^{\frac{1}{6}} \text{Hz},
\end{equation}
where $\beta$ is the inverse duration of the phase transition, $H_{*}$ and $g_{*}$ are respectively the Hubble parameter and number of relativistic degrees of freedom at a temperature $T_{*}$. The latter is usually taken to be the nucleation temperature $T_{n}$ when the bubble nucleation rate is one per Hubble volume since that is when the production of GWs from all three sources is most significant. The dimensionless gravitational wave spectrum \textit{today} obtained numerically in  \cite{Hindmarsh:2015qta} is well fitted by \cite{Caprini:2015zlo}
\begin{equation}\label{eq:GWswspectrum}
\begin{split}
    h^{2}\Omega_{sw} & = 2.65\times10^{-6} \left(\frac{H_{*}}{\beta}\right)\left(\frac{\kappa\alpha}{1+\alpha}\right)^{2}\left(\frac{100}{g_{*}}\right)^{\frac{1}{3}}\ v_{w} \ S_{sw}(f)  \\
  S_{sw}(f) & = \left(\frac{f}{f_{sw}}\right)^{3}\frac{7}{4 + 3\left(f/f_{sw}\right)^{2}},
  \end{split}
\end{equation}
where $\alpha$ is the ratio of the vacuum energy released during the phase transition to the energy of the radiation bath, and $\kappa$ is the fraction of the energy of the vacuum energy that gets converted into the bulk motion of the fluid. Aside from $\alpha$ and $\kappa$, the density is parameterized in terms of other properties of the phase transition such as the ratio of the inverse duration of the phase transition to the conformal Hubble at the phase transition $\mathcal{H}_*$, the bubble wall velocity $v_w$, and the number of relativistic degrees of freedom at $T_{*}$.

We consider only long wavelength waves, with $f < f_{\text{sw}}$.  The $f^3$ scaling is a universal property of short duration phase transition. Rather than directly calculate this power spectrum in our model, we perform a simple rescaling by
\begin{equation}
        h^{2}\Omega_{\text{gw}} = \frac{\langle \tilde{h}^2 \rangle}{\langle \tilde{h}^2\rangle_0}h^{2}\Omega_{\text{sw}} \\, 
\end{equation}
where we define the averaging procedure here by averaging over a period in $\tilde{\tau}$ of $2 \pi$ at $\tau \to \infty$, when the gravitational waves are well within the horizon and lose their sensitivity to the equation of state. We denote by $\langle \tilde{h}^2\rangle_0$ the averaged gravitational wave in a purely radiation-dominated universe.  Taking the now-arbitrary normalization $J_* = 1$, we find
\begin{equation}
\langle \tilde{h}^2\rangle_0 = \frac{\tilde{\tau}^2_*}{2}.
\end{equation}
The goal for much of the remainder of this paper is then to solve the equation \eqref{eq:gw-reformulation}, with the initial conditions $\tilde{h}(k \, \tau_*) = 0$ and $\tilde{h}^\prime(k \, \tau_*) = \tilde{\tau}_*$.

Before going into that calculation, we note on the observable spectrum in the present day universe. The gravitational waves are a bit colder than the radiation, as they decouple immediately after the phase transition. The spectrum gets scaled by a factor of $a(\tau_*)^{-4}$ and we need to convert from co-moving wave number to physical frequency. We work in a convention with $a(\tau_*) = 1$, in which case $f = k /(2 \,\pi \, a(\tau_0))$.

\section{Particle Physics Models}\label{sec:model}

\begin{table}[b]
    \centering
   \setlength{\tabcolsep}{20pt}
 \begin{tabular}{c|c|c|c}
    \hline\hline
       & Particle & Degrees of freedom & Mass (GeV)\\
    \hline
         & $\Pi^0_3$& 3& $1.95\times10^5$\\
          & & 6&$2.18\times10^5$\\
          \cline{2-4}
        & $\Pi_2^\pm$ & 8 &$6.38\times10^4$\\
         & & 8 & $1.59\times10^5$\\
          &  & 12 &$1.55\times10^5$\\
           Pions & & 4 & $7.23\times10^4$\\
           \cline{2-4}
           & $\Pi_1^\pm$ & 6 &$8.90\times10^4$\\
           \cline{2-4}
           & $\Pi_1^0$ &2 &$9.82\times10^4$\\
            & & 4&$1.07\times10^3$\\
            & & 1&5.32\\
            & & 2&0\\
            \hline
             & $\Xi_1^0$ & 6&0\\
            \cline{2-4}
            & $\Xi_1^\pm$ &4 &$2.31\times10^{-7}$\\
             & & 4&$2.81\times10^{-2}$\\
              Fermions & & 4&$6.04\times10$\\
              \cline{2-4}
            & $\Xi_2^\pm$ &8 &$2.11\times10^{-7}$\\
             & & 8&$2.48\times10^{-2}$\\
              & & 8&$1.42\times10^2$\\
              \hline
              & $A^\prime$ & 2 &0\\
              \cline{2-4}
              & $G$ & 6 &0\\
              \cline{2-4}
              Gauge bosons  &$W^{\prime\pm}$& 12 & $1.19\times10^5$\\
              \cline{2-4}
              & $Z^\prime$ & 3 &$2.09\times10^5$\\
              \hline\hline
    \end{tabular}
    \caption{Numerical spectrum of the WCSM at $f_\pi =$ 80 TeV. Additionally, there are 4 light tachyonic degrees of freedom as well.}
    \label{tab:wcsm-spectrum}
\end{table}

The Weak Confined Standard Model (WCSM) is a model in which the $\text{SU}(2)_\text{L}$ component of the electroweak force is strongly coupled and weak isospin is confined. To achieve this, a scalar is coupled to the kinetic term for the gauge bosons,
\begin{equation}
    \mathcal{L} = - \frac{\Phi}{\Lambda} W^i_{\mu\nu} W^{i \mu\nu},
\end{equation}
where $\Lambda$ is the scale of the operator.  If the scalar $\Phi$ acquires an expectation value in the early universe, the value of the weak coupling $g^\prime$ is shifted.  Since the weak coupling constant is asymptotically free, it will run to strong coupling if the nominal confinement scale $\Lambda_\text{W}$ is larger than the elecotroweak phase transition temperature $T_c \lesssim v$. In order to have Big Bang Nucleosynthesis as observed, the weak coupling should return to the SM phase at a temperature $T_{\text{SM}} \gtrsim 10~\text{MeV}$.  Further details of the WCSM can be found in Ref.\ \cite{Berger:2019yxb}.

The SM has 12 $SU(2)_\text{L}$ doublet Weyl fermions and therefore has a $SU(12)$ flavor symmetry with respect to this gauge group. This symmetry gets broken to $Sp(12)$ due to chiral symmetry breaking assumed to occur at strong $SU(2)_L$ coupling. As a result, there are 65 massless Goldstone bosons (pions) in the spectrum. The flavor symmetry is explicitly broken by both the $SU(3)_C \times U(1)_Y$ gauge couplings and the fermion Yukawa couplings.  The non-isospin gauge symmetry is spontaneously broken as $SU(3)_C \times U(1)_Y \rightarrow SU(2)_C \times U(1)_Q$. Out of 9 gauge bosons, 5 gauge bosons become massive and five pions are eaten. Out of the remaining 60 pions, 58 pions acquire masses through gauge interactions and Yukawa interactions. Four of these pion masses are tachyonic, but we expect these pions to be light after minimizing their potential.

In addition to the massive pions, there are fermionic states in the WCSM which contribute to modifying the equation of state. The left-handed fermions of the SM become mostly composite, while the right-handed states are mostly elementary. The composite fermions have masses around the confinement scale $\Lambda_\text{W} \sim 4 \pi f_\pi$. Only the elementary fermions, with masses $\ll \Lambda_\text{W}$, contribute significantly to modifying the equation of state at relevant temperatures.  

The mass spectrum of the pions and the fermions is sensitive to dimensionless operator coefficients which encode loop corrections and are expected to be $\mathcal{O}$(1) numbers. The values of these coefficients can be taken to be $\pm 1$ for the purpose of calculating the numerical spectrum of the WCSM given in Table \ref{tab:wcsm-spectrum}. 

\begin{figure}[h!]
    \centering
        \includegraphics[width=0.7\textwidth]{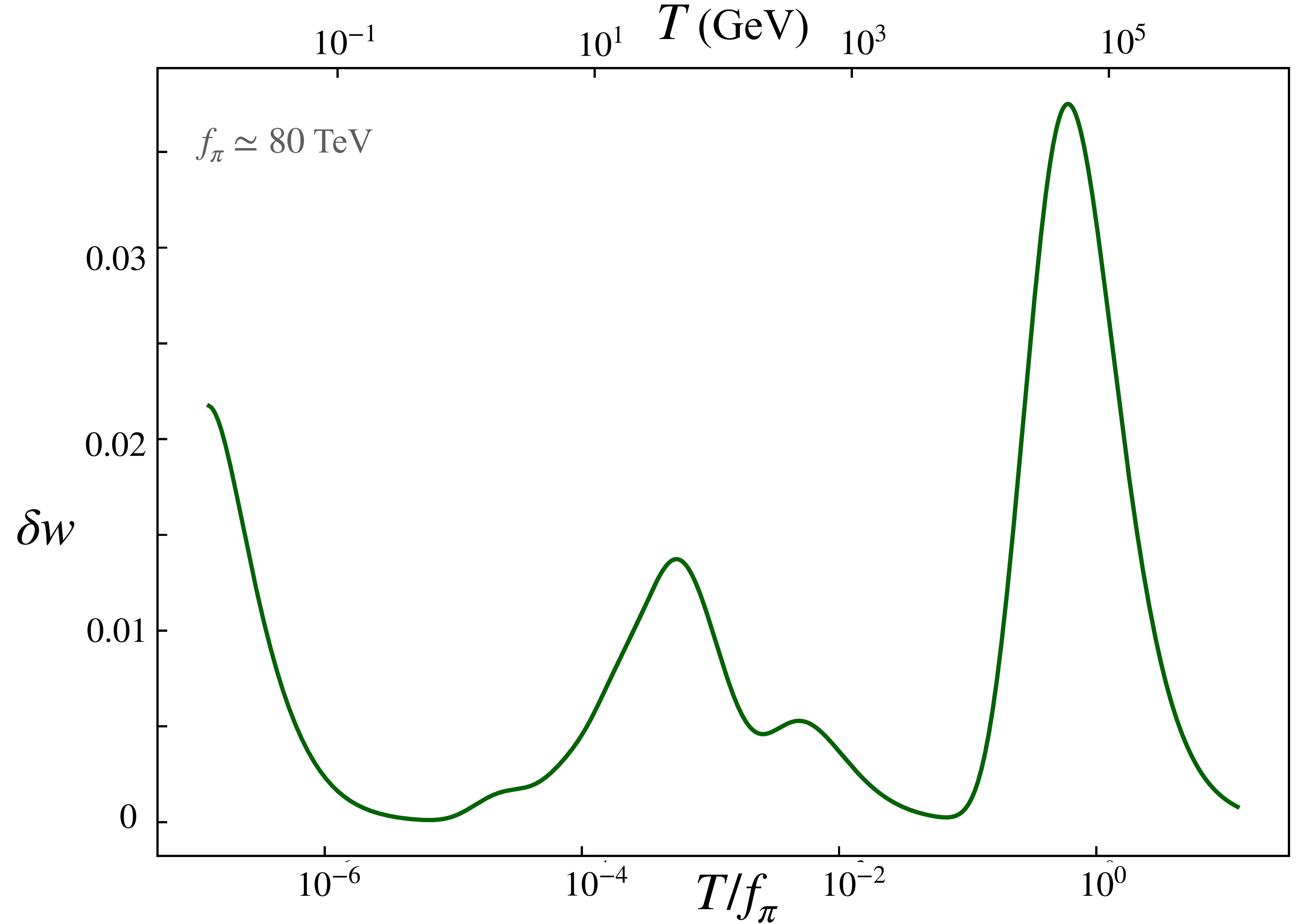}
    \caption{Deviations from $w = \frac{1}{3}$ in the WCSM for a benchmark scenario where $f_{\pi} \simeq$ 80 TeV. The peak near $T/f_{\pi} \simeq$ 1 corresponds particles of mass around $ f_\pi$ - mainly 49 pions which get their mass predominantly from the strong coupling or the top Yukawa - decoupling from the thermal bath.} \label{fig:boundsfit}
\end{figure}

With this numerical spectrum, we can repeat the procedure outlined in Sec.\ \ref{sec:eos} to calculate $\delta w$ for the WCSM. The behaviour of $\delta w$ over a range of values of $T$ for a benchmark value of $f_\pi = 80$ TeV is shown in Fig.\ \ref{fig:boundsfit}.

\begin{figure}[tbh]
     \centering
    \subfloat[][]{\label{fig:GWB11TeV}         \includegraphics[width=0.45\linewidth]{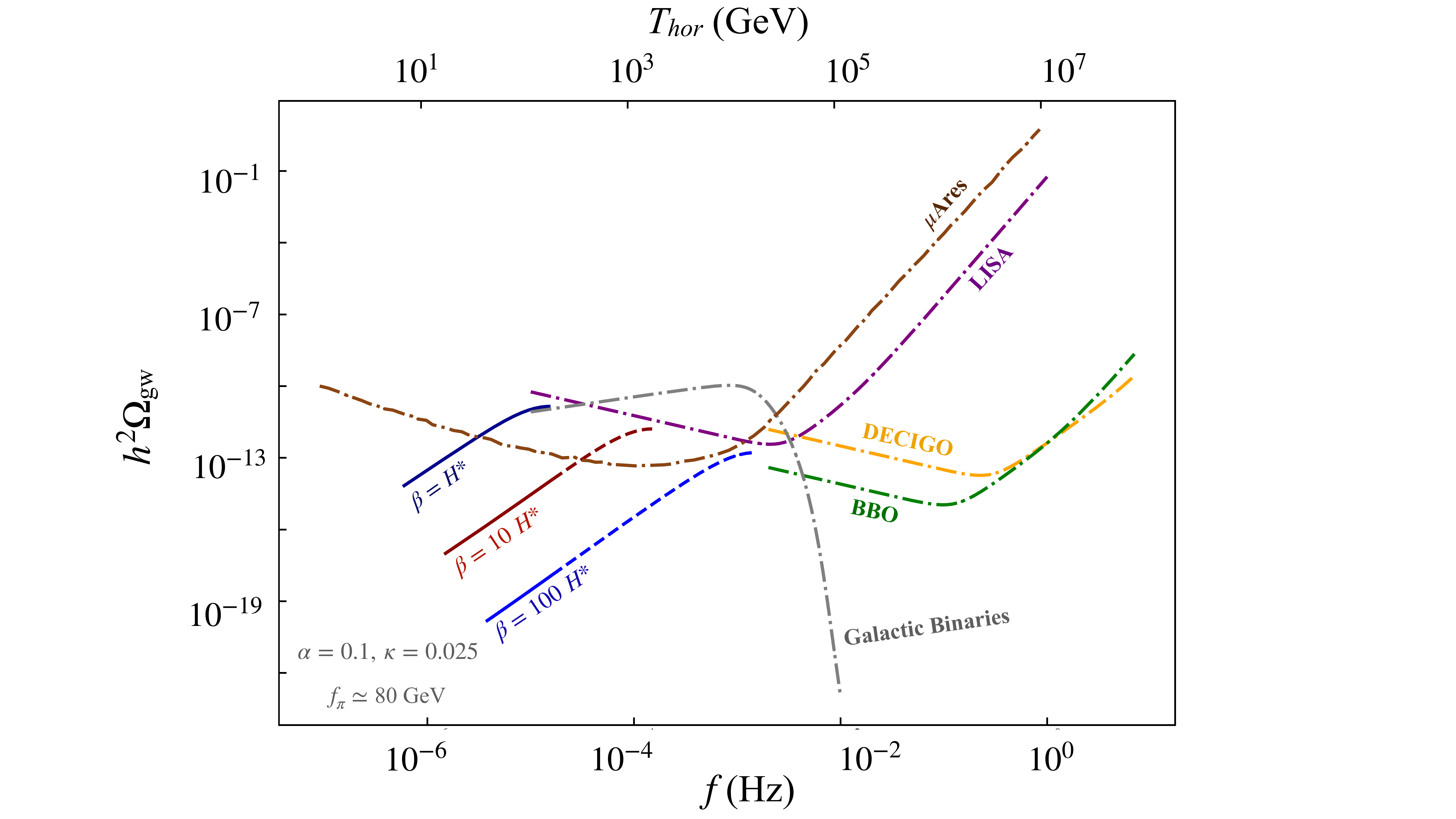}}
     \subfloat[][]{\label{fig:GWB21TeV} \includegraphics[width=0.45\linewidth]{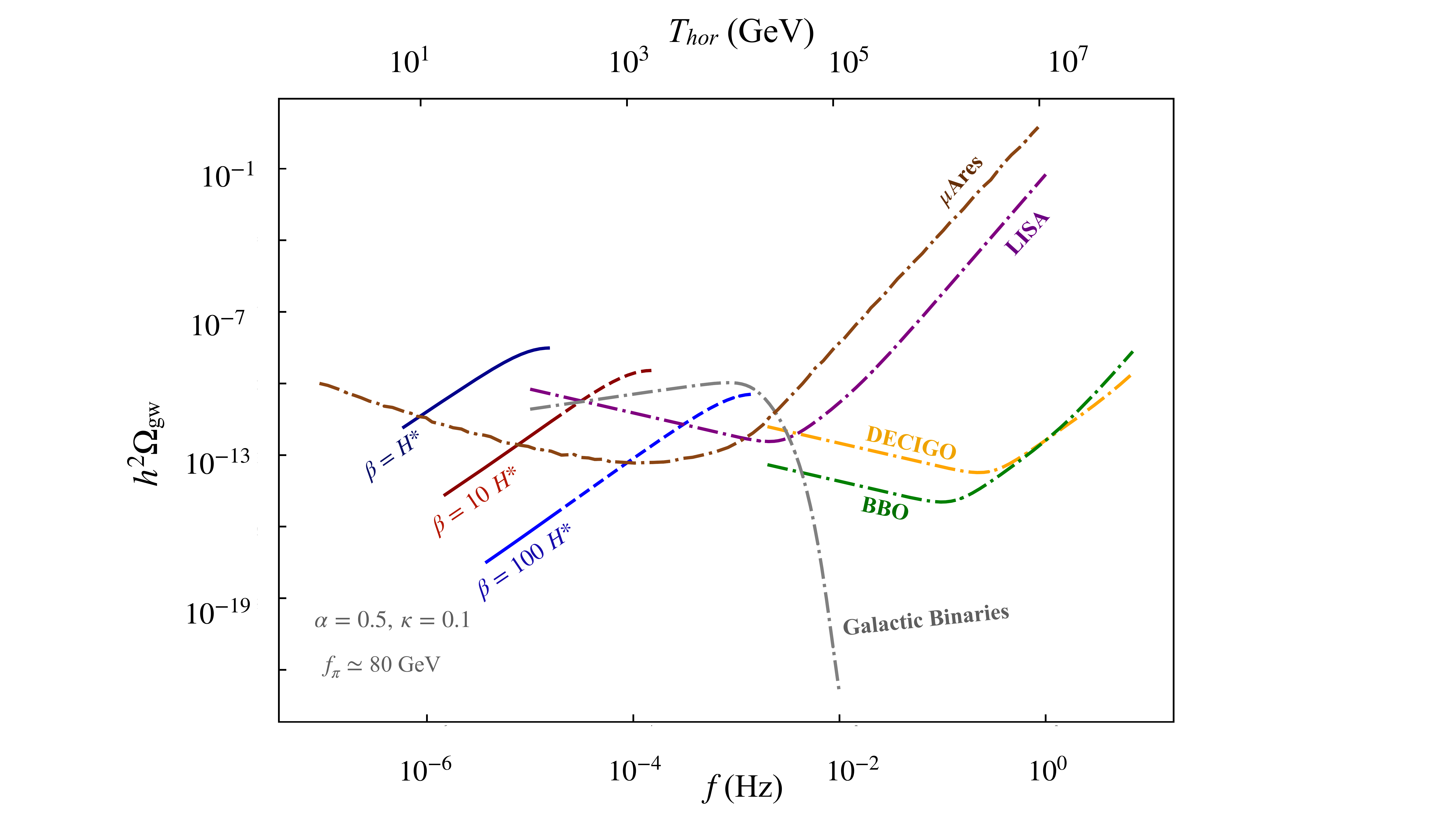}}\\
      \vspace{-0.3in}
        \subfloat[][]{\label{fig:GWB1100TeV}\includegraphics[width=0.45\linewidth]{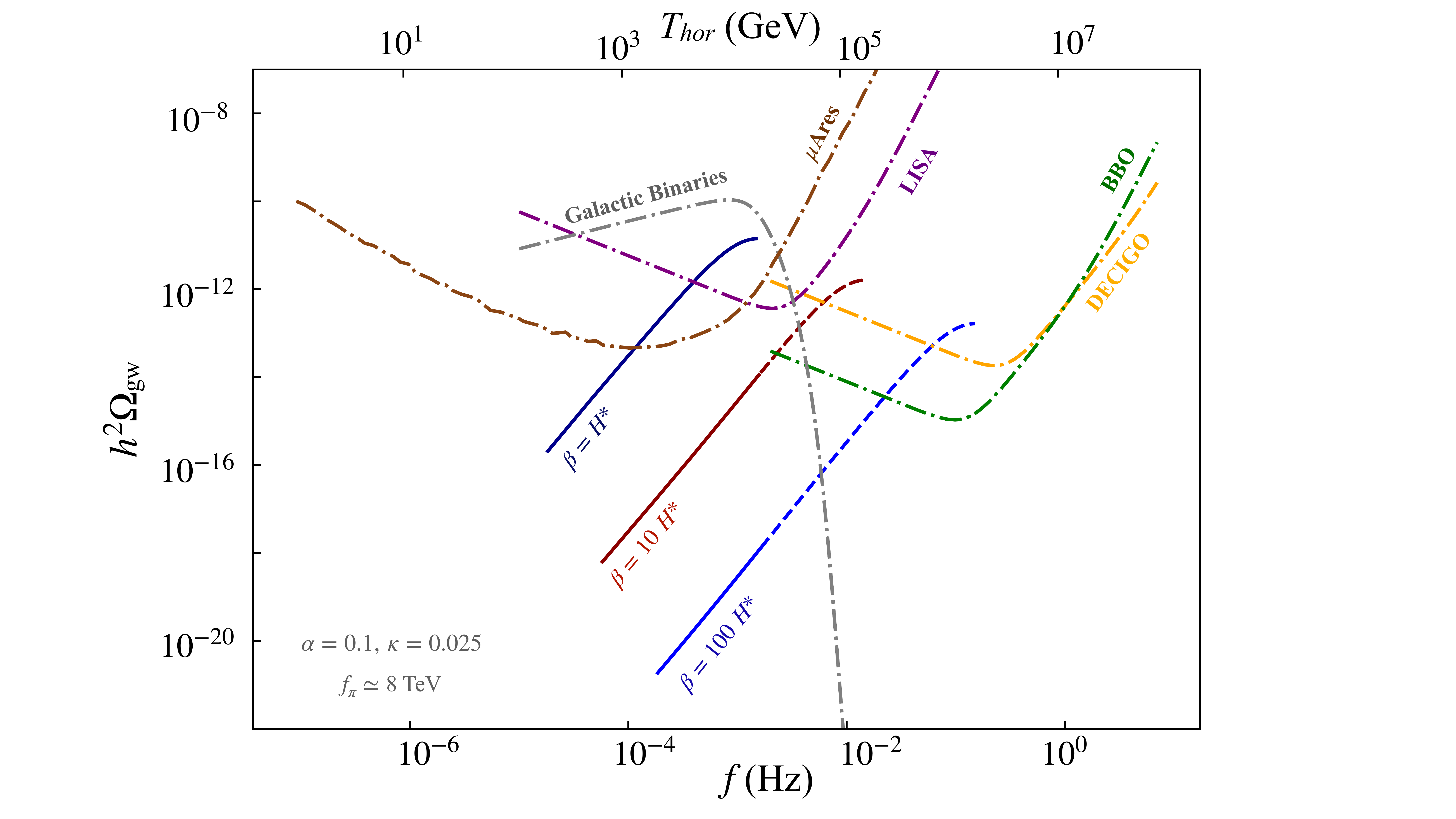}}
         \subfloat[][]{\label{fig:GWB2100TeV}\includegraphics[width=0.45\linewidth]{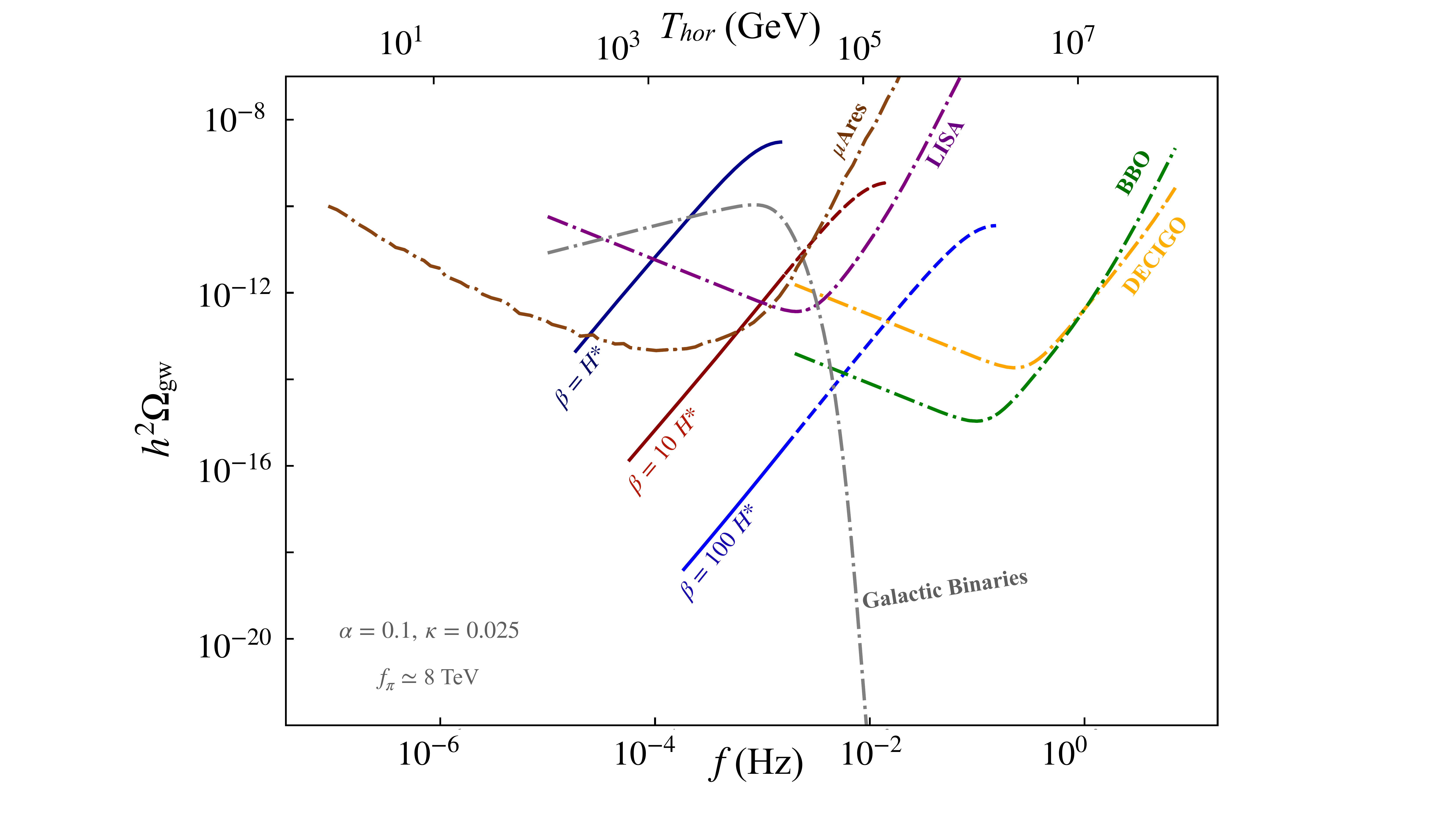}}\\
    \vspace{-0.3in}
    \subfloat[][]{\label{fig:GWB11000TeV} \includegraphics[width=0.45\linewidth]{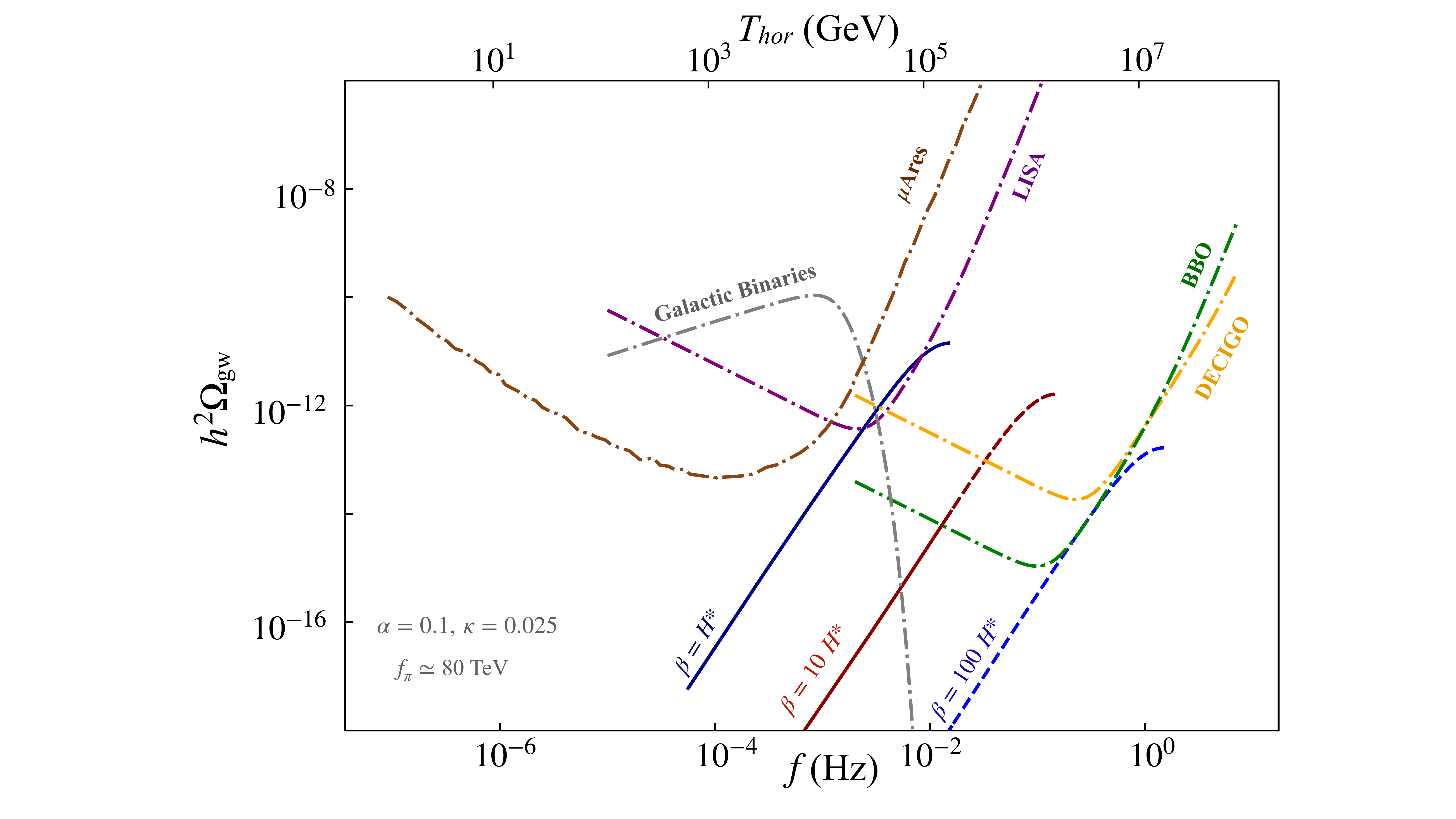}}
    \subfloat[][]{         \label{fig:GWB21000TeV} \includegraphics[width=0.45\linewidth]{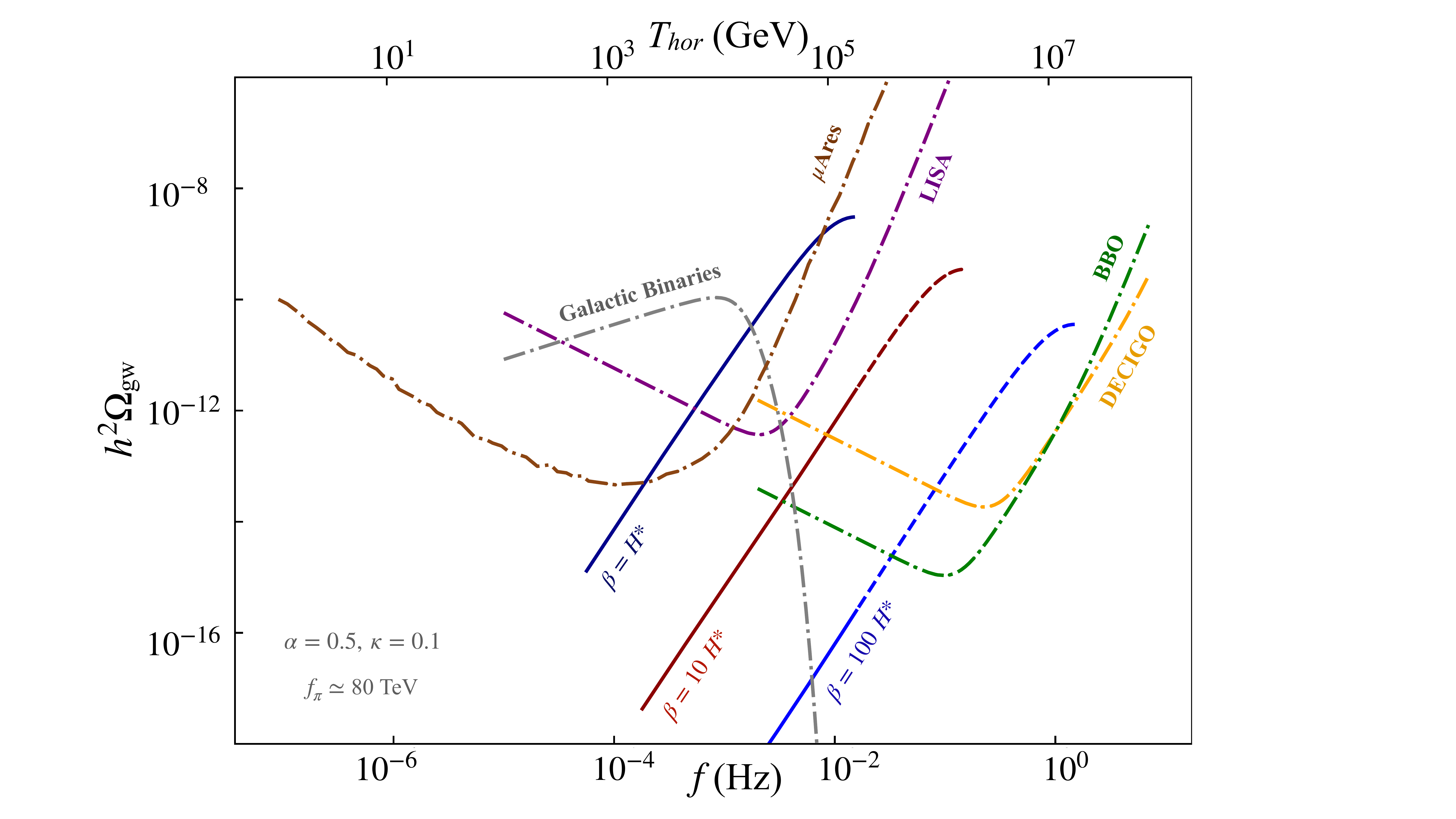}}
    \caption{The GW spectrum of an assumed first order WCSM phase transition compared to the reach of future experiments like LISA (purple), DECIGO (orange), BBO (green), and $\mu$Ares (brown). The stochastic background of GWs from galactic binaries is shown in gray. %The vertical upper horizontal axis shows the temperature of the thermal bath when a given mode enters the horizon. 
    The dotted lines in the spectrum denote frequencies with model-dependent effects. The left panels show the spectra for BI and the right panels the spectra for BII as described in the text.%Benchmark I, chosen to be fully consistent with the numerical simulation of \cite{Hindmarsh:2015qta}, is shown on the left hand column side shows for the different values of $\Lambda_{W}$ and $\beta/H_{*}$ = 1 (dark blue), 10 (dark red), and 100 (blue). Benchmark II, shown on the right hand column, extrapolates values of $\alpha$ and $\kappa$ to beyond those used in \cite{Hindmarsh:2015qta} assuming that the GW spectrum scaling is unchanged. 
    %For both benchmarks, we assume the wall velocity to be $v_{w}$ = 0.83 c, the upper limit considered.
    }
    \label{fig:SensitivityBenchmarks}
\end{figure}

For the gravitational wave spectrum, we illustrate three benchmark values for $f_\pi$ = $\left(\text{0.08, 8, 80}\right)$ \ TeV, which sets the critical temperature of the chiral phase transition $T_{C} \simeq f_{\pi}$. %$\simeq \left(\text{0.08, 8, 80 }\right)$ TeV. 
For the gravitational wave parameters in~\eqref{eq:GWswspectrum}, the precise value of $\beta/H_{*}$ is obtainable only through a numerical lattice study of the WCSM. That study is beyond the scope of this work and we therefore adopt three benchmarks, $\beta/H_{*}$ = 1, 10, 100, noting that since we are assuming that the phase transition occurs quasi-instantaneously, $\beta/H_{*}$ = 1 is a lower limit. Furthermore, we adopt two benchmark (BI and BII) values for $\alpha$ and $\kappa$. BII assumes values for the latter that are fully consistent with the numerical simulation of \cite{Hindmarsh:2015qta} while BII is an  extrapolation of the latter assuming that the GW spectrum scaling with respect $\alpha$ and $\kappa$ is unchanged. For both benchmarks, we assume the wall velocity to be $v_{w}$ = 0.83 c, the upper limit considered \cite{Hindmarsh:2015qta}. With these, we show the GW spectrum of an assumed first order WCSM phase transition and compare them to the reach of future experiments like LISA, DECIGO, and BBO. We also show the expected GW spectrum from galactic binaries, the dominant background for our model. The dotted lines in the spectrum are used to describe modes entering the horizon when the validity of the equation of state is uncertain (we describe this in more detail in the next section).

\section{Results}\label{sec:results}

We have solved \eqref{eq:gw-reformulation} numerically in the WCSM phase with initial conditions described in Section~\eqref{sec:formalism}. % which, for convenience, we reproduce here
%\begin{equation*}
%\begin{split}
%    \tilde{h}^{\prime\prime} + 2 \, \left(\tilde{H} - \frac{1}{\tilde{\tau}}\right) \, \tilde{h}^\prime & + \left(1 + \frac{2}{\tilde{\tau}} \, \left[\frac{1}{\tilde{\tau}} - \tilde{H}\right]\right) \, \tilde{h} = 0, \\
%     \tilde{h}(\tilde{\tau}_{*}) = 0 \ \ \ \ &\text{and} \ \ \ \ \tilde{h}^{\prime}(\tilde{\tau}_{*}) = \tau_{*}.
%\end{split}
%\end{equation*}
As discussed above, the particle physics effects of the WCSM on this equation describing the propagation of GWs are encoded in the small changes to the Hubble parameter due to deviations of the equation of state from $w = 1/3$ during the radiation domination era. The latter is determined assuming the validity of chiral perturbation theory, but there is some uncertainty, inherent to the strongly coupled nature of this theory, on what the critical temperature for the phase is, what the properties and nature of the phase transition are, and what distortions of the equation of state would occur due to relatively strong pion interactions close to the confinement scale. A robust answer to these questions can only come from a detailed numerical lattice study, which is beyond the scope of this work. 

As such, we trace the high frequency end of our spectrum within an order of magnitude of the cutoff scale $\Lambda_{\chi} = 4 \pi f_{\pi}$ for chiral perturbation theory, as a dashed line. Furthermore, any modes that would not quite reach a point where they lose sensitivity to the equation of state before the universe deconfines back to the SM phase are subject to modification by that deconfining phase transition and changes to the equation of state just below that phase transition. Therefore, modes for which $10 \pi / k < \tau_{\text{SM}}$, the conformal time of the transition to the SM phase, are also drawn as dashed lines.

\begin{figure}[h]
    \centering
    \includegraphics[width=0.7\textwidth]{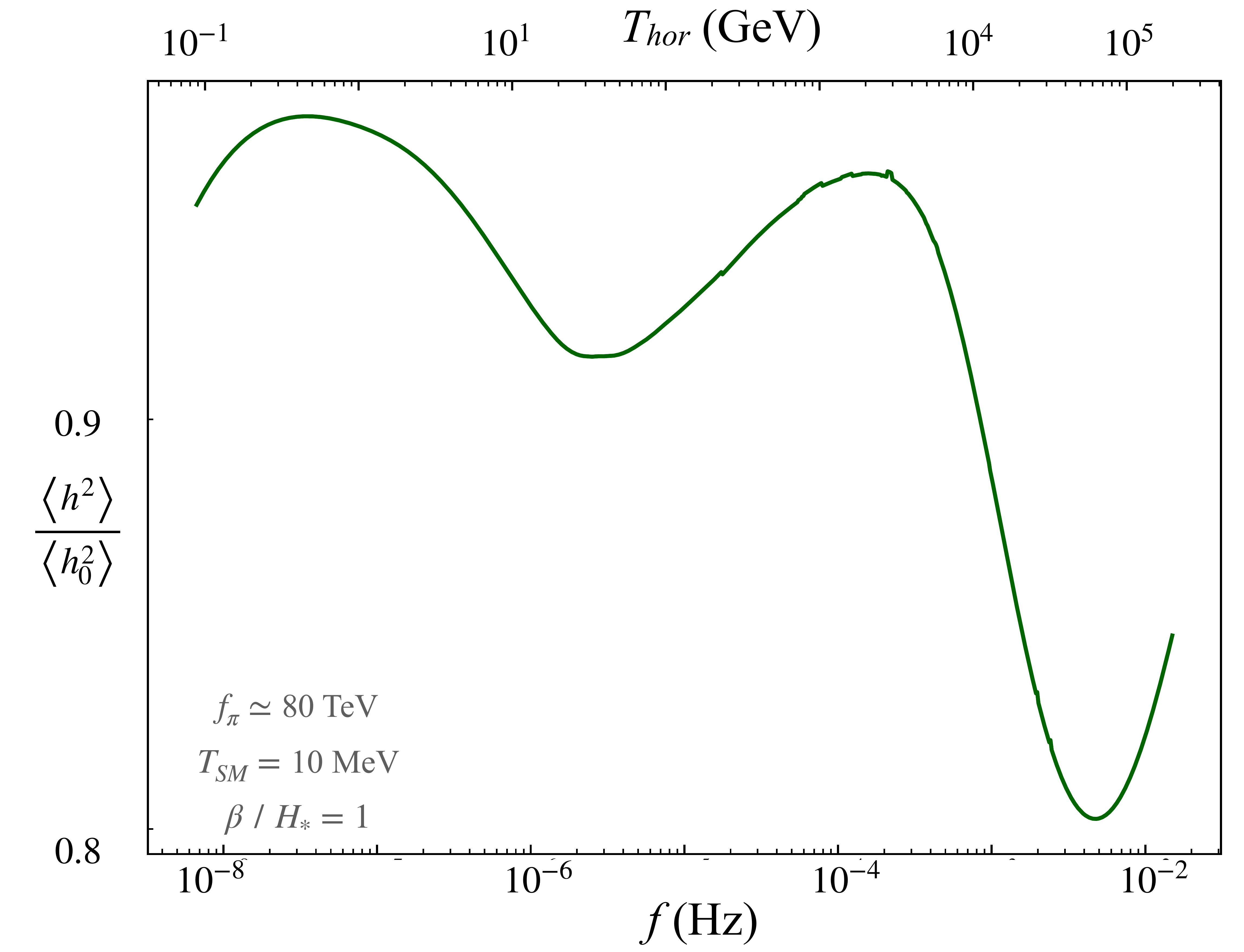}
    \caption{Average gravitational wave power relative to that in the absence of changes to the equation of state as a function of comoving wavenumber in (GeV). The vertical axis is the temperature of the thermal bath at the moment of horizon entry for each mode. The largest change - the dip at $k \lesssim$ 10$^{-7}$ GeV corresponds to the mode that enters the horizon when particles of mass $\mathcal{O}(f_\pi)$ - mainly 49 pions - decouple from the thermal bath.   }
    \label{fig:wcsm-shift}
\end{figure}

In Figure~\ref{fig:wcsm-shift}, we plot the ratio of the time-averaged gravitational wave amplitude today for $\delta w \neq$ 0 compared to the $\delta w $ = 0 case for $f_{\pi} \simeq$  80 TeV. Since this value of $f_{\pi}$ corresponds to the variation of the equation of state with respect to temperature (upper axis) shown in Figure~\ref{fig:boundsfit}, we can compare the two figures and see that the largest change occurs at $k \lesssim$ 10$^{-9}$ GeV and corresponds to the mode that enters the horizon when particles of mass around $ f_\pi$ (mainly 49 pions which get their mass predominantly from the strong coupling or the top Yukawa) become non-relativistic. 

\begin{figure}[h]
    \centering
    \includegraphics[width=0.7\textwidth]{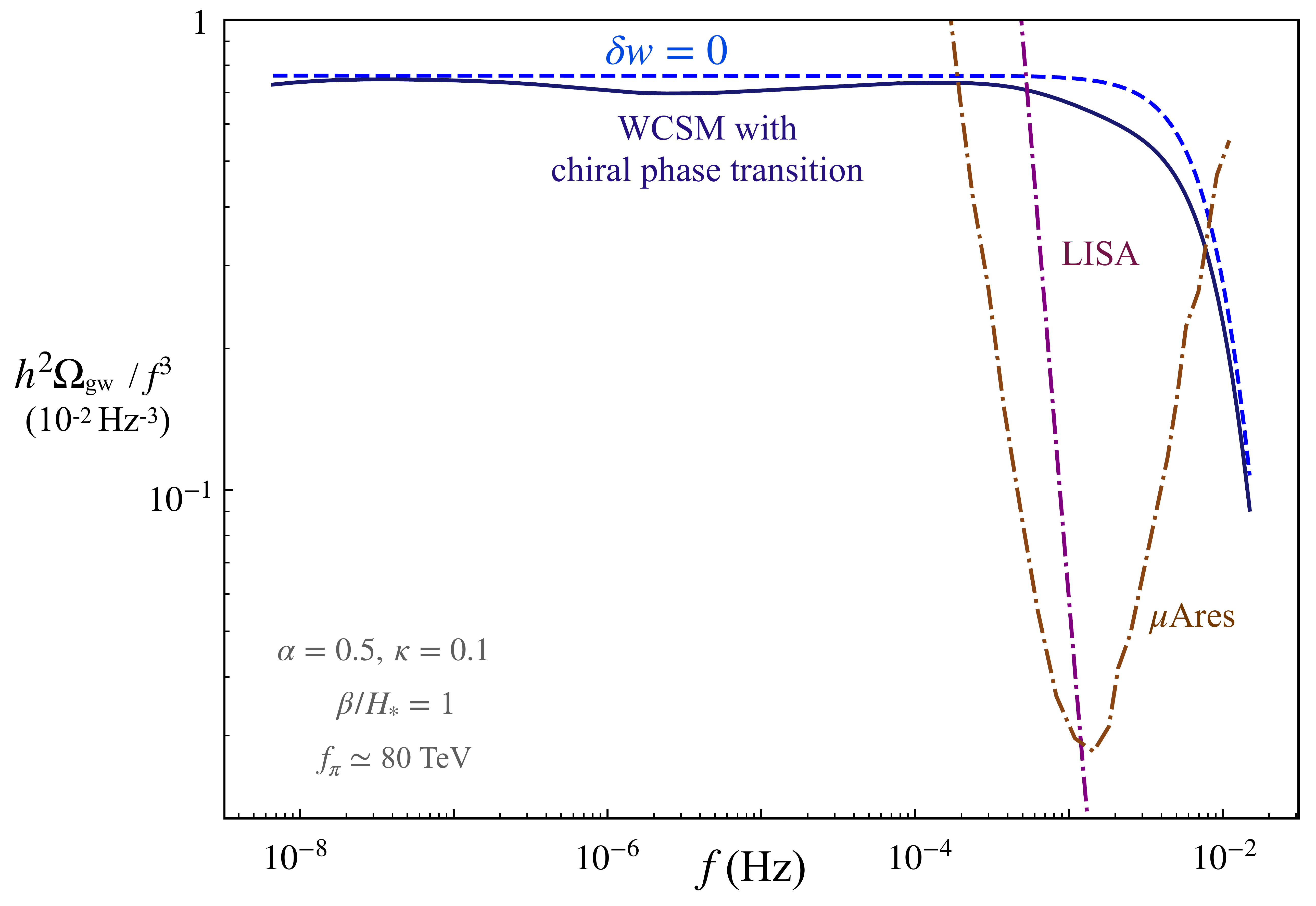}
    \caption{A zoomed in plot Figure~\ref{fig:GWB21000TeV} with only the $\beta/H_{*}$ = 1 benchmark case shown. To make the $f^{3}$ scaling at low frequencies apparent, we plot $h^{2}\Omega_\text{gw}/f^{3}$ instead of $h^{2}\Omega_\text{gw}$. The blue curve represents the $\delta w = 0$ case, corresponding to no conformal symmetry breaking, the dark blue curve represents WCSM with chiral symmetry breaking. We see that LISA and, especially, $\mu$Ares have some sensitivity to the two curves in a region where they defer and the universal $f^{3}$ scaling is valid, making such an effect potentially observable. }
    \label{fig:wcsm-lisa}
\end{figure}

We also show a zoomed in version of the  $f_\pi$ = 80 TeV, $\alpha$ = 0.5, $\kappa$ = 0.1, $\beta/H_{*}$ = 1 benchmark (Fig.\ \ref{fig:GWB21000TeV}) in Fig.\ \ref{fig:wcsm-lisa}. To make the $f^{3}$ scaling at low frequencies apparent, we plot $h^{2}\Omega_\text{gw}/f^{3}$ instead of $h^{2}\Omega_\text{gw}$. We show the $\delta w = 0$ case, corresponding to no conformal symmetry breaking, and the WCSM with chiral symmetry breaking, where conformal symmetry is strongly broken. We see that LISA has some sensitivity to the two curves in a region where the conformal and non-conformal scenarios differ which lies at the upper end of the region where the causality limited $f^{3}$ scaling is valid. There are also pronounced dips well into the region where the $f^{3}$ scaling is valid, but this lies in the $\mu\text{Hz}$ gap described in Ref. \cite{Fedderke:2021kuy}. Unfortunately, the sensitivity of the proposed detection scheme in this work still does not quite reach the sensitivity required to see the deviation from the expected $f^{3}$ scaling for this benchmark, but we hope that our model adds further motivation to thinking about future detectors capable of probing this frequency range, where there is already significant interest for astrophysical sources of GWs.   

\section{Discussion and Outlook}\label{sec:outlook}

The signal we describe here is quite general and applies to any scenario with a long wavelength gravitational wave background and a significant number of particles becoming non-relativistic.  For example, one could examine supersymmetric scenarios with a large spectrum of superpartners at similar masses.  The signal also bears some resemblance to that studied by Ref.\ \cite{Dienes:2021woi}.  One could imagine slight tweaks of that scenario to get similar dips.

One further intriguing possibility is to modify the QCD phase transition such that it is first order.  One scenario in which this would occur is a supercooled electroweak phase transition~\cite{Iso:2017uuu}, as the QCD transition would then occur with six light flavors of quarks. After the QCD phase transition, the amount of radiation in the universe is rather low. As such, the process of the muons becoming non-relativistic is sufficient to induce a roughly 5\% change in $w$, which is more challenging to observe. %Nevertheless, this 5\% dip could be observable in Pulsar Timing Array (PTA) measurements of GWs.

While the prediction of a $f^3$ scaling of the power, up to modifications due to the equation of state, is universal, the normalization of the spectrum and peak frequency are dependent on the properties of the phase transition. Further calculations are needed to determine these properties of the phase transition.  Lattice gauge theory is rather challenging for this model due to the addition of the Higgs field, but this would be the most robust way to study the phase transition.  A lattice study would also elucidate the nature of the strongly interacting phase, which we assume be chiral symmetry breaking as in QCD.  It is possible that the gravitational wave properties can be also estimated from a Nambu-Jona-Lasinio model~\cite{Aoki:2017aws,YuanyuanWang:2022nds}. 

Note that any dimensionful parameters in the model could in principle lead to such a deviation in the spectrum.  In particular a super-renormalizable scalar cubic coupling or non-renormalizable operator will lead to a small shift in the power of the low frequency spectrum.

The promising signal type presented in this work will ultimately require further study by gravitational wave telescopes.

{\bf Acknowledgments.} We thank Akshay Ghalsasi for useful discussions and Andrew Long for feedback on the manuscript. JB would like to thank the Mainz Institute for Theoretical Physics and the Aspen Center for Physics for their hospitality during the completion of parts of this work.  JB and BP are supported by the National Science Foundation under Grant No.\ 2112789. 

\nocite{*}
\bibliography{gwspectrum}

\bibliographystyle{apsrev4-1}
\end{document}